\documentclass{article}
\usepackage{spconf,amsmath,graphicx,hyperref}
\usepackage{enumitem}
\usepackage{amsmath}
\usepackage{amssymb}
\usepackage{booktabs}

\title{ROAM-ASD: Robust Open-World Active Speaker Detection with Flexible Multimodal Fusion}
\name{Pu Wang~$^{1}$ \qquad Hugo Van hamme~$^{2}$}
  
  \address{$^{1, 2}$~KU Leuven, Department of Electrical Engineering, Leuven, Belgium \\
      $^{1}$~pu.wang@esat.kuleuven.be, $^{2}$~hugo.vanhamme@kuleuven.be}
\begin{document}
\ninept
\maketitle
\begin{abstract}
Active speaker detection (ASD) requires reliable association between visible faces and acoustic speech, yet existing systems often degrade under challenging domains or incomplete observations. We introduce ROAM-ASD, a robust audiovisual framework that jointly models audio, full-face, and fine-grained mouth representations. A unified joint self-attention mechanism processes all input streams together with modality-agnostic query tokens, enabling direct interaction among available modality inputs. Modality dropout further improves robustness when input streams are unavailable. ROAM-ASD achieves state-of-the-art performance across five ASD benchmarks: $98.8\%$ mAP on WASD, $87.9\%$ on UniTalk, $96.5\%$ on AVA, $99.3\%$ on ASW, and $98.2\%$ on Talkies, improving over previous best systems by $5.1$, $4.7$, $0.9$, $1.0$, and $2.1$ mAP points, respectively. ROAM-ASD also substantially improves zero-shot cross-dataset generalization and remains robust to missing observations.
\end{abstract}
\begin{keywords}
Active speaker detection, audio-visual fusion, multimodal learning, robustness, joint self-attention
\end{keywords}
\section{Introduction}
\label{sec:intro}
Active speaker detection (ASD) aims to identify, for each visible face in a scene, whether that person is producing audible speech. It is important for many downstream applications, including speaker diarization, human-robot interaction, and multimodal conversational systems. ASD models typically address this task by jointly analyzing acoustic speech activity and speech-related facial motion, making effective audiovisual fusion central to reliable detection.

Recent years have seen substantial progress in audiovisual modeling for ASD. ASC~\cite{alcazar2020active} organizes audiovisual observations from multiple visible speakers over time and learns pairwise and temporal relations to capture long-range multi-speaker dependencies. MAAS~\cite{alcazar2021maas} represents speech events and candidate speakers in a graph, using graph reasoning to associate detected speech with the corresponding visible speaker. TalkNet~\cite{tao2021someone} performs bidirectional cross-attention between audio and visual representations, followed by self-attention to capture longer-term dependencies. ASDNet~\cite{kopuklu2021design} combines audiovisual features with inter-speaker relation modeling and temporal context, while Light-ASD~\cite{liao2023light} adopts a lightweight fusion architecture with recurrent modeling for efficient detection. LoCoNet~\cite{wang2024loconet} captures long-range dependencies within each speaker track together with short-term interactions among different speakers. Beyond architectural fusion, TalkNCE~\cite{jung2024talknce} strengthens audio-visual correspondence through talk-aware contrastive learning. TS-Talk~\cite{jiang23c_interspeech} introduces target-speaker information into the fusion process to condition detection on a specific speaker, while LASER~\cite{Nguyen_2026_WACV} incorporates lip-landmark guidance during training to emphasize localized speech-related facial motion. 
\vspace{-0.5em}
\begin{figure}[!t]
\centerline{\includegraphics[width=1.0\linewidth]{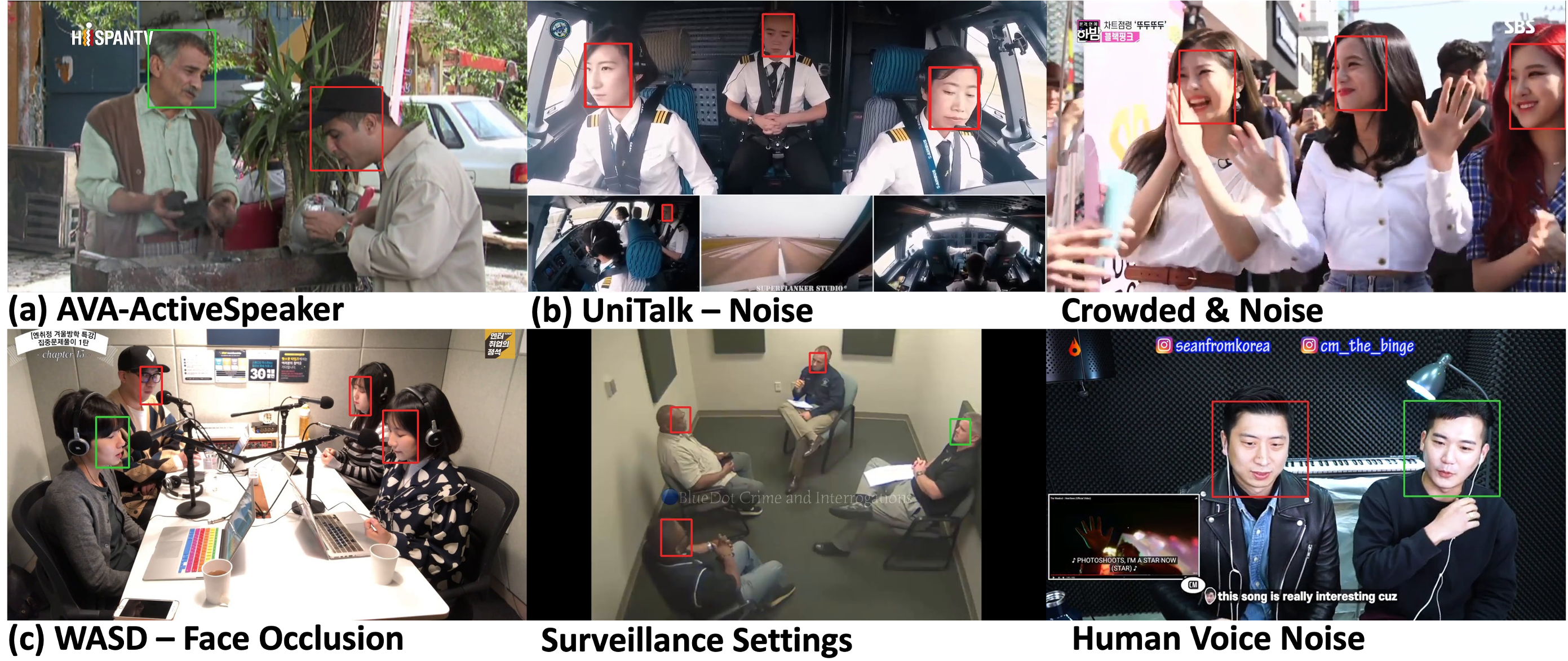}}
\vspace{-1.2em}
\caption{Example frames from (a) AVA; (b) UniTalk; and (c) WASD. Bounding boxes indicate visible speakers, with green denoting active speakers and red denoting non-speaking faces.}
\label{fig:dataset}
\vspace{-1em}
\end{figure}
\vspace{-3em}
\begin{figure}[!htbp]
\centerline{\includegraphics[width=1\linewidth]{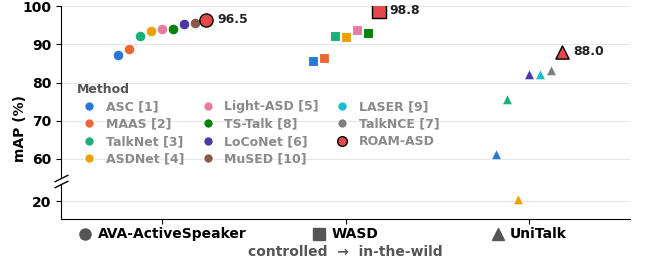}}
\vspace{-1.2em}
\caption{Previously reported mAP (\%, higher is better) of ASD methods on AVA~\cite{tao2024enhancing}, WASD~\cite{roxo2024wasd}, and UniTalk~\cite{nguyen2025revisiting}, together with
ROAM-ASD results. Each color denotes the same method across datasets, marker shapes distinguish datasets.}
\label{fig:generalmap}
\vspace{-5.5mm}
\end{figure}
\vspace{-1em}
\begin{figure*}[!htbp]
\centerline{\includegraphics[width=0.99\linewidth]{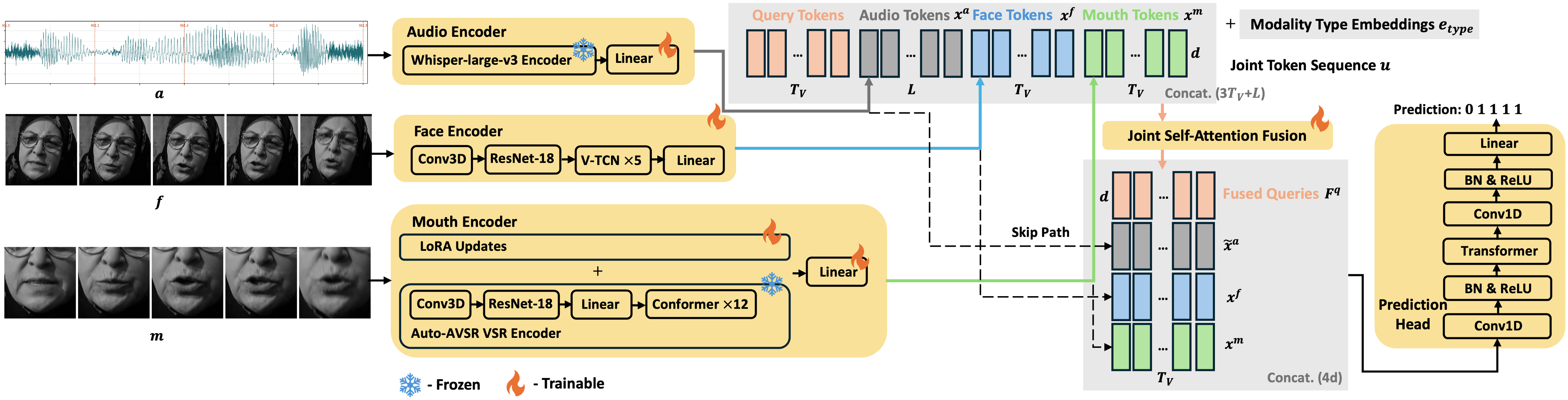}}
\vspace{-1.2em}
\caption{Overview of ROAM-ASD. The model processes audio, full-face, and mouth streams using separate encoders and projects them into a shared feature dimension. The resulting tokens, together with learnable query tokens and modality-type embeddings, are fused by joint self-attention. Skip paths preserve the original modality features for the prediction head.}
\label{fig:architecture}
\vspace{-4mm}
\end{figure*}

However, most existing ASD methods are developed and evaluated primarily on AVA-ActiveSpeaker~\cite{roth2020ava}, the most widely used benchmark in the field. AVA is constructed from movie content and typically contains relatively clear, close-up face observations, as illustrated in Fig.~\ref{fig:dataset}(a). Real-world recordings can be substantially more challenging, involving multiple nearby speakers, interfering voices and environmental noise, 
and facial occlusion.
Representative examples are shown in Fig.~\ref{fig:dataset}(b)-(c), drawn from the recent in-the-wild benchmarks WASD~\cite{roxo2024wasd} and UniTalk~\cite{nguyen2025revisiting}. Therefore, although recent methods have nearly saturated AVA, with several achieving over 95\% mean average precision (mAP), their performance drops substantially on these more challenging datasets, as shown in Fig.~\ref{fig:generalmap}, where we summarize previously reported results across AVA, WASD, and UniTalk. For example, the strongest baseline reported on UniTalk achieves only 83.2\% mAP~\cite{nguyen2025revisiting}. 

One potential reason for this performance gap under in-the-wild conditions lies in how visual speech information is represented. Most existing ASD systems represent visual evidence primarily through full-face features, which can be distracted by non-speech facial activity. For example, in Fig.~\ref{fig:dataset}(b), a non-speaking person has an open mouth in a crowded and noisy scene, which may provide misleading visual evidence. Roxo et al.~\cite{roxo2024wasd} further show that state-of-the-art (SOTA) ASD models fail when a visible person exhibits slight lip motion or expressive mouth movements while another, off-screen person is speaking, as well as under facial occlusion and interfering background voices. In contrast, speech production is most directly reflected in localized mouth and lip motion. Fine-grained mouth representations can therefore provide complementary evidence for distinguishing true articulation from unrelated facial movements. The effectiveness of lip-landmark guidance in LASER~\cite{Nguyen_2026_WACV} further supports the value of such localized cues. However, fine-grained mouth information has rarely been explicitly modeled in existing ASD systems.

A second concern is how multimodal information is fused. Cross-attention is widely used in ASD models, such as TalkNet and TS-Talk, to exchange information between audio and visual representations. However, these interactions are typically defined between specific pairs of streams, with one providing the queries and another the keys and values. 
Extending such a design to additional inputs requires specifying additional fusion paths, tying the architecture to a predefined set of streams. This can become restrictive in real-world recordings, where some inputs may be unavailable or unreliable due to facial occlusion, poor visual resolution, or corrupted audio. In such cases, a fusion mechanism that can directly operate on the available inputs, without requiring predefined pairwise interactions, would be more robust to incomplete observations.

To address these limitations, we introduce ROAM-ASD, a robust active speaker detection framework designed for in-the-wild conditions. Rather than relying solely on a global face representation, ROAM-ASD jointly models audio, full-face, and fine-grained mouth streams as complementary inputs. The mouth-region representations are cropped directly from the detected face and therefore require no additional input source or annotation.
We further introduce a unified joint self-attention fusion mechanism in which audio, face, and mouth tokens are processed together in a single sequence. Self-attention allows each token to directly interact with all available tokens, eliminating the need to define separate pairwise fusion paths between specific input streams. During training, we apply modality dropout to randomly remove individual input streams, exposing the model to different combinations of available inputs and improving robustness to missing observations at inference time.
We conduct extensive experiments across five ASD benchmarks, covering in-domain evaluation, zero-shot cross-domain transfer, and incomplete-modality conditions. ROAM-ASD achieves consistent improvements across these settings, with particularly large gains on challenging in-the-wild benchmarks.
The main contributions are:
\begin{itemize}[itemsep=0pt, topsep=0pt, partopsep=0pt, parsep=0pt, leftmargin=*]
\item We introduce ROAM-ASD, which incorporates fine-grained mouth representations alongside audio and face features and uses unified joint self-attention with modality-agnostic query tokens to fuse available inputs without predefined pairwise fusion paths.
\item ROAM-ASD achieves SOTA performance across five ASD benchmarks: 98.8\% mAP on WASD, 87.9\% on UniTalk, 96.5\% on AVA, 99.3\% on ASW, and 98.2\% on Talkies, improving over previous best systems by $\mathbf{+5.1}$, $\mathbf{+4.7}$, $\mathbf{+0.9}$, $\mathbf{+1.0}$, and $\mathbf{+2.1}$ mAP points, respectively. It also demonstrates substantially stronger zero-shot cross-dataset generalization. ASD examples are available at~\url{https://wangpuup.github.io/ROAM-ASD/}.
\item ROAM-ASD is highly robust to scattered missing observations and maintains useful performance under extended loss of individual input streams. Examples of missing-input conditions are available at \url{https://wangpuup.github.io/ROAM-ASD-modality/}.
\end{itemize}

\section{ROAM-ASD Model}
\label{sec:model}
Following the standard track-based ASD formulation~\cite{tao2021someone, wang2024loconet, tao2024enhancing}, we process one \emph{face track} at a time. Each track consists of the scene audio and a sequence of face crops corresponding to the same person, sampled at the video's native frame rate. For each video frame $t$, the model predicts a speaking probability $p_t$, with the label $y_t\in\{0,1\}$ indicating whether the tracked person is audibly speaking. 

ROAM-ASD augments this conventional audio-face input with a fine-grained mouth stream cropped directly from each tracked face. The resulting three input streams are denoted by $c\in\{a,f,m\}$ for audio, face, and mouth, respectively. Fig.~\ref{fig:architecture} illustrates the overall architecture. Each stream is processed by a modality-specific encoder and projected to a common feature space (dimension $d=256$ here), producing audio, face, and mouth representations $\mathbf{x}^a\in\mathbb{R}^{L\times d}$, $\mathbf{x}^f\in\mathbb{R}^{T_v\times d}$, and $\mathbf{x}^m\in\mathbb{R}^{T_v\times d}$, respectively. These representations are jointly processed with query tokens through self-attention. The resulting fused query representations $\mathbf{F}^q\in\mathbb{R}^{T_v\times d}$ are combined with skip features from the individual streams and passed to a temporal prediction head to produce frame-level speaking probabilities.

\textbf{Audio encoder.}
We use the frozen encoder of Whisper-large-v3~\cite{radford2023robust} to extract audio representations from the input waveform. A trainable linear projection maps them to audio features $\mathbf{x}^a_j \in \mathbb{R}^{d}$, where $j$ indexes the audio tokens.

\textbf{Face encoder.}
Following~\cite{tao2024enhancing}, we use a compact face encoder consisting of a 3D convolutional stem, a per-frame ResNet-18 backbone, a V-TCN block, and a linear projection. Each tracked face crop is resized to $122\times122$ and passed through the encoder to produce face features $\mathbf{x}^f_t \in \mathbb{R}^{d}$. The face encoder is trained from scratch jointly with the downstream ASD model.

\textbf{Mouth encoder.}
We adopt the pretrained visual speech recognition encoder from Auto-AVSR~\cite{ma2023auto}, consisting of a 3D convolutional frontend, a ResNet-18 backbone, and a Conformer encoder pretrained for lipreading on 3291 hours of visual speech data. Since the pretrained model is not optimized for ASD, we adapt it using rank-8 LoRA~\cite{hu2022lora} on the attention projections while keeping the original encoder parameters frozen. A trainable linear projection maps the adapted features to mouth representations $\mathbf{x}^m_t \in \mathbb{R}^{d}$.

To obtain the mouth input, we apply MediaPipe FaceMesh within each
tracked face to localize the mouth region, which is aligned and resized
to $88\times88$ grayscale. Short landmark-detection failures ($<1$~s) are handled by interpolating crop coordinates between neighboring valid detections, while longer failures are treated as missing observations. 

\textbf{Joint self-attention modality fusion.}
Given the encoded modality representations, we concatenate them into a joint sequence for fusion. For a track containing $T_v$ video frames, we introduce one query token for each frame, resulting in $T_v$ query tokens. All query tokens are copies of a single learned vector $\mathbf{q}\in\mathbb{R}^{d}$, broadcast along the time axis. Unlike $\mathbf{x}^a$, $\mathbf{x}^f$, or $\mathbf{x}^m$, the query tokens contain no modality-specific information and serve as modality-agnostic representations for accumulating fused information from the available input streams.
The resulting sequence contains $3T_v+L$ tokens and is initialized as $\mathbf{u}_i^{(0)}=\mathbf{z}_i+\mathbf{e}_{\mathrm{type}(i)}$, where $\mathbf{z}_i\in\{\mathbf{q},\,\mathbf{x}_j^a,\, \mathbf{x}_t^f,\,\mathbf{x}_t^m\}$ and $\mathbf{e}_{\mathrm{type}(i)}\in\mathbb{R}^{d}$ is the corresponding type embedding for audio, face, and mouth tokens, and zero for query tokens. Each token is associated with its own physical timestamp, which is incorporated into self-attention later using RoPE~\cite{su2024roformer} together with an ALiBi-style attention bias~\cite{press2022train}. This makes the initially identical query tokens frame-specific. When an input stream is unavailable, its tokens are masked out during self-attention fusion, while the query tokens and tokens from the remaining streams remain available.
The joint sequence is processed by 4 pre-norm Transformer layers with 256-dimensional, 4-head self-attention and a 1024-dimensional feed-forward network with GELU activation. Self-attention is performed jointly over the entire sequence, allowing query, audio, face, and mouth tokens to interact directly within every layer. After the final layer, the query-token outputs are layer-normalized and retained as the fused representations $\mathbf{F}^q\in\mathbb{R}^{T_v\times d}$. Since these query tokens are modality-agnostic and never masked, the fusion does not depend on any particular input stream being present and can therefore operate when one or more streams are missing.

\textbf{Skip connections and prediction head.}
The fused query representations $\mathbf{F}^q$ are combined with skip features from the individual input streams before frame-level prediction. These skip connections provide the prediction head with direct access to the original encoder representations, reducing the risk that useful frame-level cues are overly smoothed or specialized by the fusion module. Since the audio representations are produced at a different temporal rate from the video, they are linearly interpolated to the video frame positions, yielding $\tilde{\mathbf{x}}^a_t$. The face and mouth representations, $\mathbf{x}^f_t$ and $\mathbf{x}^m_t$, are already frame-aligned. For each video frame $t$, we concatenate the fused query representation with the three stream-specific features, $[\tilde{\mathbf{x}}^a_t;\mathbf{x}^f_t;\mathbf{x}^m_t;\mathbf{F}^q_t]\in\mathbb{R}^{4d}$. The concatenated features are first projected to 128 dimensions using a 1D convolution, followed by an 8-head, 128-dimensional Transformer encoder layer with a 512-dimensional feed-forward network for temporal modeling, and a 1D convolution with kernel size 3. A final linear layer maps the resulting features to two output logits, followed by a softmax to obtain the frame-level speaking probability.

\textbf{Modality dropout training.}
Although joint self-attention can operate with missing inputs, training only on complete observations may cause the model to rely heavily on all three streams. We therefore apply modality dropout during training. For each clip, the face and mouth streams are independently dropped with probability 0.15. For audio, we distinguish between missing input and observed silence: the audio stream is removed with probability 0.15, while with probability 0.10 the waveform is set to zero but the stream is retained. Dropout decisions are independent across streams, exposing the model to different combinations of available inputs.

\section{Experiments}
We evaluate ROAM-ASD on AVA~\cite{roth2020ava} and four in-the-wild
benchmarks: ASW~\cite{kim21k_interspeech},
Talkies~\cite{alcazar2021maas}, WASD~\cite{roxo2024wasd}, and
UniTalk~\cite{nguyen2025revisiting}. \textbf{AVA} contains 38.5 hours
of movie content, while \textbf{ASW} and \textbf{Talkies} contain
30.9 and 4.2 hours of in-the-wild video, respectively.
\textbf{WASD} contains 30 hours organized into five conditions:
\emph{Optimal Conditions (OC)}, \emph{Speech Impairment (SI)},
\emph{Face Occlusion (FO)}, \emph{Human Voice Noise (HVN)}, and
\emph{Surveillance Settings (SS)}. \textbf{UniTalk} contains more than
44.5 hours covering \emph{Underrepresented Languages},
\emph{Background Noise}, \emph{Crowded Scenes}, and
\emph{Mixed (Above) Conditions}.

All models follow the same training recipe. We use AdamW with a learning rate of $10^{-4}$ and weight decay of $10^{-4}$, with 500 iterations of linear warm-up followed by cosine decay to $10^{-6}$. Training uses mixed precision and gradient clipping at 5.0. For data augmentation,  we add MUSAN noise~\cite{snyder2015musan} at an SNR uniformly sampled from 5 to 20~dB. 

We conduct two sets of experiments. First, we evaluate ASD performance under complete audiovisual observations using both in-domain and zero-shot cross-domain evaluation. For in-domain evaluation, models are trained and evaluated on the corresponding dataset, whereas zero-shot evaluation directly evaluates a model trained on one dataset on another without adaptation. 
Second, we evaluate robustness to missing inputs by removing audio, face, mouth, or both visual streams. We consider both continuous and scattered missing observations. For continuous missing segments, we use holes spanning $5\%$, $15\%$, or $30\%$ of each track, repeated to reach $30\%$ total coverage, as well as single holes spanning $50\%$ or $75\%$ of the track and complete ($100\%$) stream removal.
We report mAP over the labeled frames within the missing regions and measure degradation relative to the same model evaluated on the corresponding frames without missing inputs. We compare models trained with and without modality dropout.

\section{Results}
We first evaluate in-domain ASD performance on UniTalk (Table~\ref{tab:unitalk}), WASD (Table~\ref{tab:wasd}), and AVA, Talkies, and ASW (Table~\ref{tab:ava_talkies_asw}). The best result in each setting is shown in \textbf{bold}. ROAM-ASD consistently outperforms previous methods across all five benchmarks, achieving $87.9\%$, $98.8\%$, $96.5\%$, $98.2\%$, and $99.3\%$ mAP on UniTalk, WASD, AVA, Talkies, and ASW, respectively. Compared with the strongest baseline on each benchmark, this corresponds to absolute improvements of $\mathbf{4.7}$, $\mathbf{5.1}$, $\mathbf{0.9}$, $\mathbf{2.1}$, and $\mathbf{1.0}$ mAP points, respectively. The gains are particularly pronounced on the more challenging UniTalk and WASD benchmarks, as shown in Fig.~\ref{fig:generalmap}. Representative detection examples from all five benchmarks are available at \url{https://wangpuup.github.io/ROAM-ASD/}.

We further evaluate zero-shot cross-dataset generalization, where a model trained on one dataset is directly evaluated on another without adaptation. In the UniTalk study~\cite{nguyen2025revisiting}, TalkNCE shows the strongest cross-dataset performance among the evaluated baselines, and we therefore use it as the reference baseline in Table~\ref{tab:zero_shot}. ROAM-ASD outperforms TalkNCE in all $12$ cross-dataset settings. The gains are particularly pronounced when training on ASW, with improvements of $59.3$, $28.9$, and $41.0$ mAP points on AVA, Talkies, and UniTalk, respectively. When trained on UniTalk, ROAM-ASD also improves over TalkNCE by $5.2$, $2.9$, and $8.7$ points on AVA, Talkies, and ASW, respectively. These results demonstrate substantially stronger cross-dataset generalization across different domains.

\begin{figure}[!ht]
\centerline{\includegraphics[width=1.1\linewidth]{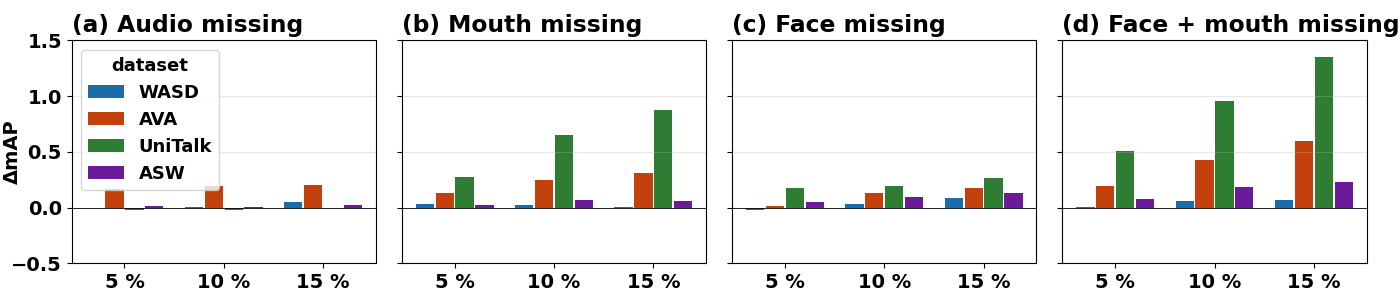}}
\vspace{-1em}
\caption{$\Delta\mathrm{mAP}= \mathrm{mAP}_{\text{no-gap}}- \mathrm{mAP}_{\text{missing}}$ under scattered missing observations at different missing ratios.}
\label{fig:sparse}
\vspace{-1.2em}
\end{figure}

\begin{figure}[!htbp]
\centerline{\includegraphics[width=1.1\linewidth]{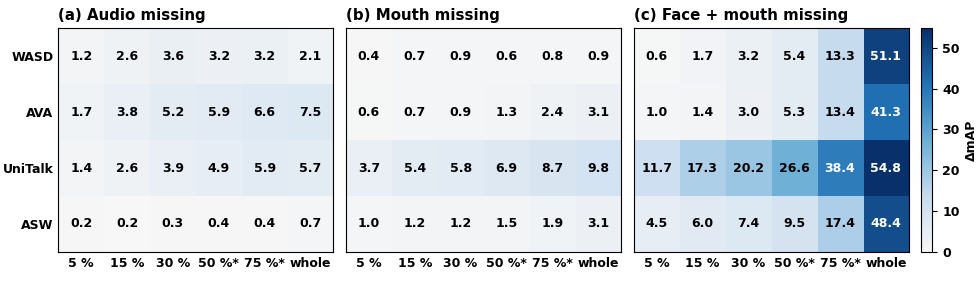}}
\vspace{-1em}
\caption{$\Delta\mathrm{mAP}= \mathrm{mAP}_{\text{no-gap}}- \mathrm{mAP}_{\text{missing}}$ under continuous missing segments at different missing ratios.}
\label{fig:relative-1}
\vspace{-1.2em}
\end{figure}

We next evaluate the robustness of ROAM-ASD to incomplete input streams at inference time. We consider two temporal patterns of missing observations: \emph{scattered} missing observations, where missing portions are distributed throughout the sequence, and \emph{continuous} missing segments, where observations are unavailable for a contiguous portion of the sequence. Fig.~\ref{fig:sparse} shows the results under scattered missing observations. ROAM-ASD is highly robust to this type of input loss. Across all four datasets and missing-stream configurations, removing up to $15\%$ of the observations causes at most approximately $1.5$ mAP points of degradation. This indicates that scattered short-term missing observations have little impact on ROAM-ASD. We next consider the more challenging case of continuous missing segments in Fig.~\ref{fig:relative-1}. Since missing the face stream alone results in little performance degradation, we omit this condition. ROAM-ASD remains relatively robust to continuous mouth loss on WASD, AVA, and ASW, with only $0.9$, $3.1$, and $3.1$ mAP points of degradation, respectively, even when the mouth stream is unavailable for the entire sequence. UniTalk is more sensitive, with a degradation of $9.8$ points. In contrast, the degradation increases substantially as both face and mouth observations become unavailable for longer periods. Under complete removal of both visual streams, this behavior follows from the nature of the ASD task itself: ASD determines whether a particular visible face is producing the observed speech. Without any visual information about the target face, audio alone can indicate the presence of speech but cannot associate that speech with the target person. Consequently, different face tracks sharing the same audio become indistinguishable to the model.

\begin{figure}[!htbp]
\centerline{\includegraphics[width=1.1\linewidth]{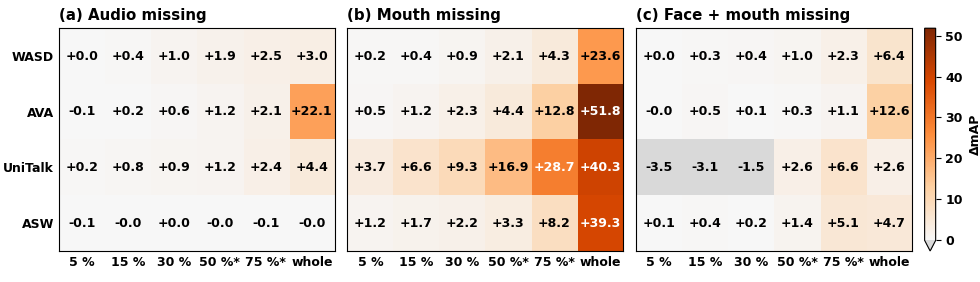}}
\vspace{-1.2em}
\caption{Modality-dropout gain ($\Delta\mathrm{mAP}_{\text{no-dropout}} - \Delta\mathrm{mAP}_{\text{ROAM-ASD}}$) under continuous missing segments. Positive values indicate reduced degradation.}
\label{fig:relative-2}
\vspace{-2em}
\end{figure}

Fig.~\ref{fig:relative-2} examines the contribution of modality-dropout training under continuous missing segments. Positive values indicate reduced degradation compared with training without modality dropout. The benefit generally becomes larger as the missing ratio increases, particularly for mouth removal. When the mouth stream is completely unavailable, modality dropout reduces the degradation by $23.6$, $51.8$, $40.3$, and $39.3$ mAP points on WASD, AVA, UniTalk, and ASW, respectively. These results show that modality dropout is particularly important when an input stream is unavailable for an extended period. Representative detection examples under missing-input conditions are available at \url{https://wangpuup.github.io/ROAM-ASD-modality/}.

\begin{table}[!htpb]
\vspace{-1.5em}
\centering
\caption{mAP ($\uparrow$, \%) on UniTalk under in-domain and zero-shot.}
\label{tab:unitalk}
\resizebox{0.49\textwidth}{!}{
\setlength{\tabcolsep}{4pt}
\begin{tabular}{lcccccc}
\toprule
& \multicolumn{5}{c}{In-domain} & Zero-shot \\
\cmidrule(lr){2-6} \cmidrule(lr){7-7}
Model & Overall & Language & Crowded & Noise & Mixed
      & AVA$\rightarrow$UniTalk \\
\midrule
ASDNet~\cite{kopuklu2021design}   & $20.6$ & $30.8$ & $17.5$ & $14.8$ & $20.3$ & -- \\
ASC~\cite{alcazar2020active}      & $61.4$ & $74.7$ & $62.9$ & $53.4$ & $57.3$ & -- \\
TalkNet~\cite{tao2021someone}  & $75.7$ & $80.1$ & $77.6$ & $67.1$ & $70.3$ & -- \\
LoCoNet~\cite{wang2024loconet}  & $82.2$ & $85.8$ & $84.6$ & $80.0$ & $76.2$ & -- \\
LASER~\cite{Nguyen_2026_WACV}   & $82.2$ & $86.7$ & $83.7$ & $81.6$ & $75.8$ & -- \\
TalkNCE~\cite{jung2024talknce}  & $83.2$ & $86.7$ & $84.9$ & $84.1$ & $77.9$ & $77.5$ \\
\midrule
ROAM-ASD & $\mathbf{87.9}$ & $\mathbf{91.6}$ & $\mathbf{89.3}$
         & $\mathbf{90.6}$ & $\mathbf{81.2}$ & $\mathbf{82.6}$ \\
\bottomrule
\end{tabular}
}
\vspace{-1.5em}
\end{table}

\begin{table}[!htbp]
\vspace{-1em}
\centering
\caption{Zero-shot cross-dataset mAP ($\uparrow$, \%) comparison. Each cell reports
TalkNCE / ROAM-ASD.}
\label{tab:zero_shot}
\resizebox{0.48\textwidth}{!}{
\setlength{\tabcolsep}{8pt}
\begin{tabular}{lcccc}
\toprule
Train $\backslash$ Eval & AVA & Talkies & ASW & UniTalk \\
\midrule
AVA
& $95.5 / \mathbf{96.5}$ & $88.3 / \mathbf{88.4}$ & $88.5 / \mathbf{98.3}$
& $77.5 / \mathbf{82.6}$ \\

Talkies
& $55.7 / \mathbf{60.9}$ & $95.6 / \mathbf{98.2}$
& $84.5 / \mathbf{96.4}$ & $59.9 / \mathbf{65.4}$ \\

ASW
& $29.2 / \mathbf{88.5}$ & $58.8 / \mathbf{87.7}$
& $96.1 / \mathbf{99.3}$ & $33.8 / \mathbf{74.8}$ \\

UniTalk
& $88.0 / \mathbf{93.2}$ & $91.4 / \mathbf{94.3}$
& $90.4 / \mathbf{99.1}$ & $83.2 / \mathbf{87.9}$ \\
\bottomrule
\end{tabular}
}
\vspace{-1.5em}
\end{table}

\begin{table}[!htbp]
\centering
\caption{mAP ($\uparrow$, \%) on WASD under in-domain and zero-shot.}
\label{tab:wasd}
\resizebox{0.48\textwidth}{!}{
\setlength{\tabcolsep}{4pt}
\begin{tabular}{lccccccc}
\toprule
& \multicolumn{6}{c}{In-domain} & Zero-shot \\
\cmidrule(lr){2-7} \cmidrule(lr){8-8}
Model & OC & SI & FO & HVN & SS & Overall & AVA$\rightarrow$WASD \\
\midrule
ASC~\cite{alcazar2020active}       & $91.2$ & $92.3$ & $87.1$ & $66.8$ & $72.2$ & $85.7$ & $74.6$ \\
MAAS~\cite{alcazar2021maas}      & $90.7$ & $92.6$ & $87.0$ & $67.0$ & $76.5$ & $86.4$ & $70.7$ \\
ASDNet~\cite{kopuklu2021design}    & $96.5$ & $97.4$ & $92.1$ & $77.4$ & $77.8$ & $92.0$ & $79.2$ \\
TalkNet~\cite{tao2021someone}   & $95.8$ & $97.5$ & $93.1$ & $81.4$ & $77.5$ & $92.3$ & $85.0$ \\
TS-Talk~\cite{jiang23c_interspeech}   & $96.8$ & $97.9$ & $94.4$ & $84.0$ & $79.3$ & $93.1$ & $85.7$ \\
Light-ASD~\cite{liao2023light} & $97.8$ & $98.3$ & $95.4$ & $84.7$ & $77.9$ & $93.7$ & $86.2$ \\
\midrule
ROAM-ASD  & $\mathbf{99.6}$ & $\mathbf{99.7}$ & $\mathbf{98.9}$
          & $\mathbf{96.3}$ & $\mathbf{96.2}$ & $\mathbf{98.8}$
          & $\mathbf{92.7}$ \\
\bottomrule
\end{tabular}
}
\vspace{-1.5em}
\end{table}

\begin{table}[!htbp]
\centering
\caption{mAP ($\uparrow$, \%) comparison on AVA, Talkies, and ASW.}
\label{tab:ava_talkies_asw}
\resizebox{0.49\textwidth}{!}{
\begin{tabular}{lccccc}
\toprule
Dataset & EASEE~\cite{alcazar2022end} & Light-ASD~\cite{liao2023light} & LoCoNet~\cite{wang2024loconet} & MuSED~\cite{tao2024enhancing} & ROAM-ASD \\
\midrule
AVA     & 94.1 & 94.1 & 95.2 & 95.6 & $\mathbf{96.5}$ \\
Talkies & 93.6 & 93.9 & 96.1 & --   & $\mathbf{98.2}$ \\
ASW     & --   & --   & --   & 98.3 & $\mathbf{99.3}$ \\
\bottomrule
\end{tabular}
}
\vspace{-1em}
\end{table}

\section{Conclusion}
\label{sec:conclusion}
We presented ROAM-ASD, a robust ASD framework that jointly models audio, full-face, and fine-grained mouth information. ROAM-ASD uses unified joint self-attention with modality-agnostic query tokens to fuse all available inputs without predefined pairwise fusion paths, while modality dropout exposes the model to incomplete inputs during training. Experiments across five ASD benchmarks demonstrate consistent improvements over previous methods in both in-domain and zero-shot cross-dataset evaluation, with particularly strong gains on the challenging WASD and UniTalk benchmarks. ROAM-ASD is also highly robust to scattered missing observations and maintains useful performance under extended loss of individual input streams. Our analysis further shows that modality dropout becomes increasingly important as missing intervals grow. 

\newpage
\section*{Acknowledgment}
The authors thank Xiaomi Corporation, Beijing, China for their support in pursuing this research.
\bibliographystyle{IEEEbib}
\bibliography{strings,refs}

\end{document}